\documentclass[a4paper,11pt]{article}

\usepackage{hyperref}
\usepackage{amsmath,amsthm,amssymb}
\usepackage{xcolor}
\usepackage{graphicx}
\usepackage{overpic}
\usepackage{bbold}
\usepackage{textcomp}
\usepackage{siunitx}
\usepackage{physics}

\usepackage{tikz}

\usepackage{pgfplots}
\usepackage{authblk}

\begin{document}

\title{Large Scale Finite-Element Simulation of Micromagnetic Thermal Noise}

\author[1]{Florian Bruckner \thanks{florian.bruckner@univie.ac.at}}
\author[2]{Massimiliano d'Aquino}
\author[3]{Claudio Serpico}
\author[1]{Claas Abert}
\author[1]{Christoph Vogler}
\author[1]{Dieter Suess}

\affil[1]{Christian Doppler Laboratory of Advanced Magnetic Sensing and Materials, Faculty of Physics, University of Vienna, Austria}
\affil[2]{Department of Technology, University of Napoli "Parthenope", Italy}
\affil[3]{Department of Electrical Engineering, University of Napoli "Federico II", Italy}

\maketitle

\begin{abstract}
An efficient method for the calculation of ferromagnetic resonant modes of magnetic structures is presented. Finite-element discretization allows flexible geometries and location dependent material parameters. The resonant modes can be used for a semi-analytical calculation of the power spectral density of the thermal white-noise, which is relevant for many sensor applications. The proposed method is validated by comparing the noise spectrum of a nano-disk with time-domain simulations.
\end{abstract}

\section{Introduction}
The determination of micromagnetic resonant modes of magnetic nano-structures is of great importance for applications like spin-torque oscillators or magnetic sensors. The resonant modes are correlated with the thermal magnetic white-noise contributions, which may be the dominant noise source in sensor applications within sensor applications with frequencies $f \gtrsim \SI{1}{kHz}$ \cite{suess_topologically_2017}.

Ferromagnetic-resonance measurements (FMR) can be performed experimentally or numerical methods can be used to calculate the corresponding oscillation modes. Time-domain micromagnetic simulations with an applied deterministic field, like harmonic oscillation or a field pulse, etc. can be used and Fourier analysis of the magnetization response allows to determine the resonant modes \cite{grimsditch_magnetic_2004,mcmichael_magnetic_2005,roy_micromagnetic_2010}. The deterministic excitations allows to derive the oscillation frequency, but the resulting mode amplitudes will depend on the detailed form of the excitation. This problem can be overcome by using stochastic excitations which correctly describe the thermodynamic equilibrium \cite{zhu_thermal_2002}. However, required time steps in the range of femtoseconds limit the applicability of this method. Solving the problem in the frequency-domain provides an elegant and efficient way for the semi-analytic calculation of the resonant modes \cite{albert_domain_2016,baker_proposal_2017}.

This work presents a generalization of the algorithms based on a finite-difference (FD) discretization \cite{daquino_computation_2008,daquino_novel_2009} and finite-element (FEM) discretization \cite{forestiere_finite_2009}. The method is based on a finite-element micromagnetic code \cite{abert_magnum_2013}, which ensures efficient calculations for large-scale applications with spatially varying material parameters.

Furthermore, the analytical calculation of the resulting noise power spectral density \cite{bertram_experimental_2002,bertram_thermal_2002,jin_quasi-analytical_2004} is demonstrated and perfectly agrees with results calculated numerically from Langevin dynamic simulations.

\section{Eigenmode Equation}
The following calculations are based on the work of d'Aquino et al.~\cite{daquino_novel_2009}, with some modifications needed for the FEM discretization. Linearization of the Landau-Lifshitz-Gilbert equation(LLG) for small fluctuations $\mathbf{v}(\mathbf{x}, t)$ around a stable equilibrium $\mathbf{m}_0$ leads to
\begin{align} \label{eqn:linearized_LLG}
-\alpha \, \dot{\mathbf{v}} - \mathbf{m}_0 \times \dot{\mathbf{v}} &= \underbrace{\left( \mathbb{1} - \mathbf{m}_0 \otimes \mathbf{m}_0 \right)}_{\mathcal{P}_0} \, \underbrace{\gamma \, \left( h_0 \, \mathbb{1} - \mathbf{h}^\text{lin} \right)}_{\mathcal{A}_0}[\mathbf{v}],
\end{align}
with the reduced gyromagnetic ratio $\gamma=2.2127615 \cdot 10^5 \si{\metre / \ampere\second}$, the dimensionless damping parameter $\alpha$, the tangent-plane projection operator $\mathcal{P}_0$, the linearized effective field operator $\mathbf{h}^\text{lin}[\mathbf{v}] = \frac{\delta \mathbf{h}^\text{eff}}{\delta \mathbf{m}}[\mathbf{v}]$, and parallel component of the equilibrium field $h_0 = \mathbf{m}_0 \cdot \mathbf{h}^\text{eff}[\mathbf{m}_0]$ (see Appendix~\ref{sec:linearized_LLG} for a detailed derivation).

Using the Ansatz $\mathbf{v}(\mathbf{x}, t) = \sum_k a_k \, \tilde{\boldsymbol{\varphi}}'_k(\mathbf{x}) \, e^{i \omega'_k t}$ with the damped eigenvectors $\tilde{\boldsymbol{\varphi}}'_k$ and eigenfrequencies $\omega'_k$, allows to transform the differential equation into an algebraic equation. Due to the linear independence of the basis functions $e^{i \omega'_k t}$ each components $(\omega'_k, \tilde{\boldsymbol{\varphi}}'_k)$ needs to fulfill the following algebraic generalized eigenvalue equation:
\begin{align} \label{eqn:eigenvalue_problem_continuum}
\omega'_k \, \underbrace{(-i \alpha \, \tilde{\boldsymbol{\varphi}}'_k)}_{i \, \delta \mathcal{B} [\tilde{\boldsymbol{\varphi}}'_k]} + \omega'_k \, \underbrace{(-i \mathbf{m}_0 \times \tilde{\boldsymbol{\varphi}}'_k)}_{\mathcal{B}_0 [\tilde{\boldsymbol{\varphi}}'_k]} = \underbrace{\mathcal{P}_0 \, \mathcal{A}_0 [\tilde{\boldsymbol{\varphi}}'_k]}_{\mathcal{A}_{0\perp} [\tilde{\boldsymbol{\varphi}}'_k]}
\end{align}

\subsection{Without Damping $\alpha = 0$}
Setting $\alpha = 0$ in Eqn.~\eqref{eqn:eigenvalue_problem_continuum} leads to the following Hermitian generalized eigenvalue problem
\begin{align} \label{eqn:eigenvalue_problem_unperturbated}
\mathcal{A}_{0\perp} \, [\tilde{\boldsymbol{\varphi}}_k] = \omega_k \, \mathcal{B}_0 \, [\tilde{\boldsymbol{\varphi}}_k],
\end{align}
where $\tilde{\boldsymbol{\varphi}}_k$ and $\omega_k$ represent the undamped eigenmodes and eigenfrequencies, respectively. Both operators $\mathcal{A}_{0\perp}$ and $\mathcal{B}_0$ are Hermitian, which allows using the Lanczos algorithm for the solution of the eigenvalue problem, resulting in more stable and efficient calculations. Furthermore it leads to the following orthogonality relation for the undamped eigenmodes (see \cite{daquino_novel_2009} for a detailed prove) which will simplify the calculation of the PSD:

\begin{align} \label{eqn:orthogonality}
(\tilde{\boldsymbol{\varphi}}_h, \tilde{\boldsymbol{\varphi}}_k)_{\mathcal{A}_{0\perp}} = \omega_h \, (\tilde{\boldsymbol{\varphi}}_h, \tilde{\boldsymbol{\varphi}}_k)_{\mathcal{B}_0} = \delta_{hk},
\end{align}
where $(\tilde{\boldsymbol{\varphi}}_h, \tilde{\boldsymbol{\varphi}}_k)_\mathcal{O}$ denotes the complex inner product with respect to the operator $\mathcal{O}$.

Note that although the solutions of the generalized eigenvalue problem are complex, the final result for $\mathbf{v}(\mathbf{x},t)$ is real. One can show that if $(\omega_k, \tilde{\boldsymbol{\varphi}}_k)$ is a solution then so is $(-\omega_k, \tilde{\boldsymbol{\varphi}}_k^*)$ \cite{daquino_novel_2009}. If one further assumes equal amplitudes $a_k$ for each pair of eigenmodes the resulting fluctuations are real:
\begin{align}
\begin{split}
\mathbf{v}(\mathbf{x},t) &= \sum_{k/2} a_k \, \left[\tilde{\boldsymbol{\varphi}}_k(\mathbf{x}) \, e^{i \omega_k t} + \tilde{\boldsymbol{\varphi}}_k^*(\mathbf{x}) \, e^{-i \omega_k t} \right] \\
              &= \sum_{k/2} a_k \, \big[\Re\left\{\tilde{\boldsymbol{\varphi}}_k\right\} \, \underbrace{\left(e^{i \omega_k t} + e^{-i \omega_k t}\right)}_{\cos(\omega_k t)} + \Im\left\{\tilde{\boldsymbol{\varphi}}_k\right\} \, \underbrace{i \, \left(e^{i \omega_k t} - e^{-i \omega_k t}\right)}_{\sin(\omega_k t)} \big],
\end{split}
\end{align}
where the summation over $k/2$ means that only eigenmodes with positive eigenvalue are considered.

\subsection{Perturbation analysis for damping $\alpha > 0$}
Starting from Eqn.~\eqref{eqn:eigenvalue_problem_continuum} allowing for small damping $\alpha > 0$ and assuming perturbed quantities $\tilde{\boldsymbol{\varphi}}'_k = \tilde{\boldsymbol{\varphi}}_k + \delta\tilde{\boldsymbol{\varphi}}_k$, $\omega'_k = \omega_k + i \, \delta \omega_k$, and $\mathcal{B}'_0 = \mathcal{B}_0 + i \, \delta \mathcal{B}$ results in the following perturbated eigenvalue problem
\begin{align}
\mathcal{A}_{0\perp} \, \left[ \tilde{\boldsymbol{\varphi}}_k + \delta \tilde{\boldsymbol{\varphi}}_k \right] = \left( \omega_k + i \, \delta \omega_k \right) \, \left( \mathcal{B}_0 + i \, \delta \mathcal{B} \right) \, \left[ \tilde{\boldsymbol{\varphi}}_k + \delta \tilde{\boldsymbol{\varphi}}_k \right],
\end{align}
with the perturbed operator $\delta \mathcal{B} = - \alpha \mathbb{1}$. Direct solution of the damped eigenvalue problem can be achieved by means of the Arnoldi method, which also works for non-hermitian eigenvalue problems. However this would influence the solver performance and also destroy orthogonality of the eigenmodes.

Alternatively the perturbation of the eigenvector can be represented using the unperturbed eigenvectors $\delta \tilde{\boldsymbol{\varphi}}_h = \sum_k c_{hk} \tilde{\boldsymbol{\varphi}}_k$ and a perturbation analysis can be performed. Considering only terms up to first-order perturbations yields
\begin{align} \label{eqn:eigenvalue_problem_alpha}
\mathcal{A}_{0\perp} \, [\delta \tilde{\boldsymbol{\varphi}}_k] = \omega_k \, \mathcal{B}_0 \, [\delta \tilde{\boldsymbol{\varphi}}_k] + i \, \omega_k \, \delta \mathcal{B} \, [\tilde{\boldsymbol{\varphi}}_k] + i \, \delta \omega_k \, \mathcal{B}_0 \, [\tilde{\boldsymbol{\varphi}}_k]
\end{align}
Scalar multiplying both sides of equation \eqref{eqn:eigenvalue_problem_alpha} with $\tilde{\boldsymbol{\varphi}}_k$ and utilizing the orthogonality relation \eqref{eqn:orthogonality} allows to express the pertubation of the eigenfrequency as
\begin{align}
\delta \omega_k = -\omega_k \frac{(\tilde{\boldsymbol{\varphi}}_k, \tilde{\boldsymbol{\varphi}}_k)_{\delta \mathcal{B}}}{(\tilde{\boldsymbol{\varphi}}_k, \tilde{\boldsymbol{\varphi}}_k)_{\mathcal{B}_0}} = \omega_k^2 \, (\tilde{\boldsymbol{\varphi}}_k, \alpha \, \tilde{\boldsymbol{\varphi}}_k)
\end{align}
Finally the perturbation analysis for small damping results in an additional imaginary contribution to the eigenfrequency, which leads to a damped harmonic oscillation within the time domain. Ignoring the small perturbations of the eigenvectors $\tilde{\boldsymbol{\varphi}}_k$ one ends up with
\begin{align}
\mathbf{v}_\alpha(\mathbf{x}, t) &\approx \sum_k a_k \, \tilde{\boldsymbol{\varphi}}_k(\mathbf{x}) \, e^{i \omega_k t} \, e^{- \delta \omega_k t}
\end{align}

\subsection{FEM Discretization}
For the FEM formulation the functions $\mathbf{m}_0$, $\mathbf{h}^\text{eff}$, $\tilde{\mathbf{v}}(x)$, $\tilde{\boldsymbol{\varphi}}_k(x)$ are discretized using Lagrange $\mathcal{P}_1$ elements. Furthermore the continuous equation \eqref{eqn:eigenvalue_problem_unperturbated} needs to be applied to a FEM function $\tilde{\boldsymbol{\varphi}}_k(x) = \sum \tilde{\boldsymbol{\varphi}}_{k,j} \, \Lambda_j(x) = \underline{\tilde{\boldsymbol{\varphi}}}_k \cdot \underline{\Lambda}$, weighted by test-functions $\Lambda_i(x)$ and integrated over the domain $\Omega_\text{m}$. As motivated in Appendix~\ref{sec:FEM_discretization} (which also includes more details about the discretization of the effective field) both sides of the continuous equation are multiplied with $J_s$ which results in the following Galerkin discretization
\begin{align}
\int J_s \, \Lambda_i \, \mathcal{B}_0[\Lambda_j] \, \text{d}\Omega_\text{m} \; \tilde{\boldsymbol{\varphi}}_{k,j} = \int J_s \, \Lambda_i \, \mathcal{A}_{0\perp}[\Lambda_j] \, \text{d}\Omega_\text{m} \; \tilde{\boldsymbol{\varphi}}_{k,j},
\end{align}
where the index $k$ describes the eigenmode index, whereas the index $j$ describes the node of the FEM discretization. The FEM discretization results in the following matrix representations of the operators $\mathcal{A}_0$ and $\mathcal{B}_0$
\begin{align} \label{eqn:FEM_discretization}
\begin{split}
\underline{B}_0 = \int J_s \, \Lambda_i \, \mathcal{B}_0[\Lambda_j] \, \text{d}\Omega_\text{m} &= -i \int J_s \, \Lambda_i \, \mathbf{m}_0 \times \Lambda_j \, \text{d}\Omega_\text{m} \\
\underline{A}_0 = \int J_s \, \Lambda_i \, \mathcal{A}_0[\Lambda_j] \, \text{d}\Omega_\text{m} &= \gamma \, \int J_s \, h_0 \, \Lambda_i \, \Lambda_j \, \text{d}\Omega_\text{m} \\
&-\gamma \, \int J_s \, \Lambda_i \, \mathbf{h}^\text{lin}[\Lambda_j] \, \text{d}\Omega_\text{m},
\end{split}
\end{align}
and finally yields the discretized form of the generalized eigenvalue problem~\eqref{eqn:eigenvalue_problem_unperturbated}
\begin{align} \label{eqn:eigenvalue_problem_matrix}
\underline{A}_{0\perp} \, \underline{\tilde{\boldsymbol{\varphi}}}_k = \omega_k \, \underline{B}_0 \, \underline{\tilde{\boldsymbol{\varphi}}}_k,
\end{align}
with $\underline{A}_{0\perp} = \underline{P}_0 \, \underline{A}_0$ and the vertex-wise projection operator $\underline{P}_0 = \underline{\mathbb{1}} - \underline{\mathbf{m}}_0 \otimes \underline{\mathbf{m}}_0$.

Note that the matrix $\underline{B}_0$ is not unitary, due to the integration over the domain, but due to the Galerkin discretization $\underline{A}_{0\perp}$ and $\underline{B}_0$ are both Hermitian and $\underline{A}_{0\perp}$ is positive definite. These properties allow the usage of efficient solvers and preserve the orthogonality property of the eigenvectors:
\begin{align} \label{eqn:orthogonality_FEM}
(\underline{\tilde{\boldsymbol{\varphi}}}_h, \underline{\tilde{\boldsymbol{\varphi}}}_k)_{\underline{A}_{0\perp}} = \omega_h \, (\underline{\tilde{\boldsymbol{\varphi}}}_h, \underline{\tilde{\boldsymbol{\varphi}}}_k)_{\underline{B}_0} = \delta_{hk},
\end{align}
where $(\underline{\tilde{\boldsymbol{\varphi}}}_h, \underline{\tilde{\boldsymbol{\varphi}}}_k)_{\underline{O}} = \underline{\tilde{\boldsymbol{\varphi}}}_k^* \cdot \underline{O} \cdot \underline{\tilde{\boldsymbol{\varphi}}}_k$ denotes the complex inner product with respect to the operator $\underline{O}$.

Note that for the FEM discretization the restriction to the tangent-plane can only be exactly fulfilled on each vertex and not for each point in space. Introducing a rotated coordinate frame on each vertex with one axis parallel to $\mathbf{m}_0$ and eliminating all parallel contributions allows an efficient implementation and reduces the system size from $3N \times 3N$ to $2N \times 2N$. 

The following $3N \times 2N$ rotation-projection matrix performs the rotation into the tangent-plane and eliminates all parallel components
\begin{align}
\underline{R} = \begin{pmatrix}
                   \mathbf{e}_1 \cdot \mathbf{e}_x & \mathbf{e}_2 \cdot \mathbf{e}_x \\
                   \mathbf{e}_1 \cdot \mathbf{e}_y & \mathbf{e}_2 \cdot \mathbf{e}_y \\
                   \mathbf{e}_1 \cdot \mathbf{e}_z & \mathbf{e}_2 \cdot \mathbf{e}_z
                \end{pmatrix} && \mathbf{e}_3 = \mathbf{m}_0 && \mathbf{e}_2 = \mathbf{e}_3 \times \mathbf{e}_\text{ref} && \mathbf{e}_1 = \mathbf{e}_2 \times \mathbf{e}_3,
\end{align}
where $\mathbf{e}_x$, $\mathbf{e}_y$, $\mathbf{e}_z$ are cartesian unit-vectors, $\mathbf{e}_\text{ref}$ is an arbitrary reference direction, and $N$ is the number of mesh-vertices. Note that numerical instabilities may occur if $\mathbf{e}_\text{ref}$ is nearly parallel to $\mathbf{m}_0$.

The transformation into the rotating frame preserves the symmetry and definiteness of the original problem and finally yields
\begin{align} \label{eqn:eigenvalue_problem_projected}
\begin{split}
(\underline{R}^T \underline{A}_0 \, \underline{R}) \, (\underline{R}^T \underline{\tilde{\boldsymbol{\varphi}}}_k) &= \omega \, (\underline{R}^T \underline{B}_0 \, \underline{R}) \, (\underline{R}^T \underline{\tilde{\boldsymbol{\varphi}}}_k) \\
\underline{A}_{0\perp}^{''} \, \underline{\tilde{\boldsymbol{\psi}}}_k &= \omega \, \underline{B}_0^{''} \, \underline{\tilde{\boldsymbol{\psi}}}_k,
\end{split}
\end{align}
where the projected eigenmodes $\underline{\tilde{\boldsymbol{\psi}}}_k = \underline{R}^T \underline{\tilde{\boldsymbol{\varphi}}}_k$ and the reduced matrices $\underline{A}_{0\perp}^{''}$, $\underline{B}_0^{''}$ have been introduced. Note that the un-projected operator $\underline{A}_0$ can be used since the projection is directly fulfilled by the rotation-projection matrix $\underline{R}$.

For the effective solution of the eigenvalue problem assembling of the $\underline{A}_{0\perp}^{''}$ matrix has to be avoided. Instead all components need to be implemented as operators and iterative algorithms like the Lanczos method \cite{lehoucq_arpack_1998} needs to be used to solve for the $N$ smallest eigenvalues. Additionally the use of a customized preconditioner (compare e.g. \cite{suess_time_2002}) for the inversion of $\underline{A}_{0\perp}^{''}$ (which is necessary when calculating the smallest instead of the largest eigenvalues), can have a tremendous impact on the total performance and stability of the algorithm.

For the perturbation analysis of the damped discrete system the perturbation operator $\delta \mathcal{B}$ needs to be discretized, which leads to the following weighted mass matrix (again one multiplies with $J_s$)
\begin{align}
\delta \underline{B} = -\int J_s \, \alpha \, \Lambda_i \, \Lambda_j \, \text{d}\Omega_m
\end{align}

Using the same line of reasoning as in the continuum problem the discrete perturbation of the eigenfrequency can be expressed as
\begin{align}
\delta \omega_k = -\omega_k \frac{(\underline{\tilde{\boldsymbol{\varphi}}}_k, \underline{\tilde{\boldsymbol{\varphi}}}_k)_{\delta \underline{B}}}{(\underline{\tilde{\boldsymbol{\varphi}}}_k, \underline{\tilde{\boldsymbol{\varphi}}}_k)_{\underline{B}_0}} = -\omega_k^2 \, (\underline{\tilde{\boldsymbol{\varphi}}}_k, \underline{\tilde{\boldsymbol{\varphi}}}_k)_{\delta \underline{B}}
\end{align}
Note that result is a generalization of the original calculation \cite{daquino_novel_2009} which uses a finite-difference discretization (where the mass matrix is equal to the identity) as well as a constant $\alpha$.

\section{Thermal Noise Calculation}
In the previous section the homogeneous solution of the linearized LLG equation has been derived, which can be used to solve initial value problems. However for calculation of the thermal noise of a sensor, it is assumed that the deterministic dynamics already reached a stationary state. Thus the homogeneous solution will not contribute to the calculated sensor noise. Nevertheless the previous results are of great importance, since the eigenmode decomposition also diagonalizes the inhomogenous equations system and provides an efficient solution.

For the calculation of the noise spectrum, the sensor is treated as a linear time-invariant system (LTI), which allows all calculations to be performed in the frequency domain by using a standard Fourier transform. A thermal white-noise field $\tilde{\mathbf{h}}^\text{th}$ is applied and leads to an excitation of the individual eigenmodes. The calculation of the transfer function $H(\omega)$ allows to derive the output signal $\tilde{s}(\omega) = H(\omega) \, \tilde{\mathbf{h}}^\text{th}(\omega)$ as well as the corresponding power spectral density $S_{ss}(\omega)$. 

\subsection{Power Spectral Density (PSD)}
The power spectral density $S_{xx}(\omega)$ of a signal $x$ is defined as the power of a signal at a certain frequency. For signals with finite energy the standard Fourier transform can be used to decompose the signal into its energy spectral density. If, however the signal energy is infinite (as for example for an harmonic oscillation or thermal white-noise), a truncated Fourier transform needs to be used to calculate the power spectral density
\begin{align} \label{eqn:PSD}
S_{xx}(\omega) = \lim_{T\to\infty} \frac{1}{T} \, \left\langle \left\vert \int_0^T x(t) \, e^{-i \omega t} \, dt \right\vert^2 \right\rangle
\end{align}
where $\langle . \rangle$ means the expectation value.

By means of the Wiener-Khinchin theorem the PSD can also be expressed as the Fourier transform of the autocorrelation function $S_{xx}(\omega) = \mathcal{F} \left\{ R_{xx}(\tau) \right\}$ (which will prove useful for the use with stochastic variables like the thermal noise field $\mathbf{h}^\text{th}$). The general definition of the autocorrelation function, also suitable for stochastic variables, is given by
\begin{align}
R_{xx}(\tau) = \langle x(t+\tau) \, x(t) \rangle,
\end{align}

Furthermore for any LTI system it can be shown that the autocorrelation and the PSD of the output signal $s$ can be expressed as
\begin{align} \label{eqn:PSD_LTI_system}
\begin{split}
R_{ss}(\tau) &= h(\tau) * R_{hh}(\tau) * h(-\tau) \\
S_{ss}(\omega) &= H(\omega) \cdot S_{hh}(\omega) \cdot H^\dag(\omega),
\end{split}
\end{align}
where $h(\tau)$ and $H(\omega)$ are the transfer functions in the time- or frequency domain, respectively. $R_{hh}$ and $S_{hh}$ are the autocorrelation and the PSD of the input signal $\mathbf{h}^\text{th}$. $H^\dag$ denotes the Hermitian conjugate of $H$ and $*$ the convolution operator.

Thermal white-noise $\mathbf{h}^\text{th}$ can be represented by a random variable with zero mean and a variance which follows from the fluctuation-dissipation theorem
\begin{align}
\begin{split}
\langle \mathbf{h}^\text{th}(\mathbf{x}, t) \rangle &= 0 \\
\underbrace{\langle \mathbf{h}^\text{th}(\mathbf{x}, t) \; \mathbf{h}^\text{th}(\mathbf{x}', t') \rangle}_{R_{hh}(t' - t)} &= \frac{2 \alpha k_B T}{\gamma J_s} \, \delta(\mathbf{x}-\mathbf{x}') \, \delta(t-t'),
\end{split}
\end{align}
which leads to the commonly known constant PSD by performing a Fourier transform
\begin{align}
S_{hh}(\omega) = \frac{2 \alpha k_B T}{\gamma J_s} \, \delta(\mathbf{x}-\mathbf{x}')
\end{align}

\subsection{Transfer function $H(\omega)$}
Each LTI system is characterized by its transfer function in the frequency domain. The transfer function of the magnetic system relates the input field $\tilde{\mathbf{h}}^\text{th}$ with the output signal $\tilde{s} = H(\omega) \, \tilde{\mathbf{h}}^\text{th}(\omega)$. The applied thermal field $\tilde{\mathbf{h}}^\text{th}$ yields the magnetic fluctuations $\tilde{\mathbf{v}}^\text{th}$, which are in turn related to the output signal $\tilde{s}$, e.g. by means of the giant magnetoresistive (GMR) effect. 

For spacially varying magnetization the GMR effect can be described via a micromagnetic model including spin-diffusion \cite{abert_self_2016}. For sake of simplicity one can assume that the GMR effect can be approximated by the angle between the spacially averaged magnetization and a given reference direction $\hat{\mathbf{e}}_\text{ref}$. Thus the output signal $\tilde{s}$ is chosen as the average magnetization into the reference direction
\begin{align} \label{eqn:sensor_output}
\tilde{s}(\omega) = \frac{\int \tilde{\mathbf{v}}^\text{th} \cdot \hat{\mathbf{e}}_\text{ref} \; \text{d}\Omega_m}{\int \text{d}\Omega_m},
\end{align}
where $\tilde{\mathbf{v}}^\text{th}$ are the thermally induced magnetization fluctuations. 

These thermally induced magnetization fluctuations are solutions of the inhomogeneous LLG equations, where the thermal field $\tilde{\mathbf{h}}^\text{th}$ is added as source term. For sake of simplicity the thermal field is added to the undamped LLG~\eqref{eqn:eigenvalue_problem_unperturbated} and the perturbations due to damping are considered afterwards. The inhomogenous LLG equation within the time domain reads like
\begin{align} \label{eqn:linearized_LLG_thermal}
-\mathbf{m}_0 \times \dot{\mathbf{v}}^\text{th}(\mathbf{x}, t) &= \mathcal{A}_{0\perp}[\mathbf{v}^\text{th}(\mathbf{x}, t)] - \gamma \, \mathcal{P}_0 \, \mathbf{h}^\text{th}(\mathbf{x}, t)
\end{align}

Performing a temporal Fourier transform $\tilde{\mathbf{v}}^\text{th}(\mathbf{x}, \omega) = \int \mathbf{v}^\text{th}(\mathbf{x}, t) \, e^{i\omega t} \, dt$ leads to
\begin{align} \label{eqn:eigenvalue_problem_thermal}
\omega \, \mathcal{B}_0 \, [\tilde{\mathbf{v}}^\text{th}(\mathbf{x}, \omega)] = \mathcal{A}_{0\perp}[\tilde{\mathbf{v}}^\text{th}(\mathbf{x}, \omega)] - \gamma \, \mathcal{P}_0 \, \tilde{\mathbf{h}}^\text{th}(\mathbf{x}, \omega)
\end{align}

In order to utilise the orthogonality relation \eqref{eqn:orthogonality}, the inhomogenous solution $\tilde{\mathbf{v}}^\text{th}$ is expressed by the unperturbed eigenvectors
\begin{align} \label{eqn:v_th_decomposition}
\tilde{\mathbf{v}}^\text{th}(\mathbf{x}, \omega) = \sum \tilde{a}_k(\omega) \, \tilde{\boldsymbol{\varphi}}_k(\mathbf{x})
\end{align}

Scalar multiplying both sides of equation \eqref{eqn:eigenvalue_problem_thermal} with $\tilde{\boldsymbol{\varphi}}_h$ and utilizing the orthogonality relation allows to calculate the mode amplitudes $\tilde{a}_k(\omega)$
\begin{align}
\omega \, \underbrace{\left(\tilde{\boldsymbol{\varphi}}_h, \sum \tilde{a}_k \, \tilde{\boldsymbol{\varphi}}_k\right)_{\mathcal{B}_0}}_{\frac{1}{\omega_h} \tilde{a}_h} &= \underbrace{\left(\tilde{\boldsymbol{\varphi}}_h, \sum \tilde{a}_k \, \tilde{\boldsymbol{\varphi}}_k\right)_{\mathcal{A}_{0\perp}}}_{\tilde{a}_h} - \, \gamma \, \left(\tilde{\boldsymbol{\varphi}}_h, \mathcal{P}_0 \, \tilde{\mathbf{h}}^\text{th} \right),
\end{align}
which finally results in the following expression for the mode amplitudes
\begin{align}
\tilde{a}_h = - \frac{\gamma \, \omega_h}{\omega - \omega_h - i \, \delta \omega_h} \, \left(\tilde{\boldsymbol{\varphi}}_h, \mathcal{P}_0 \, \tilde{\mathbf{h}}^\text{th} \right),
\end{align}
where the previously calculated damped eigenfrequencies $\omega_h + i \, \delta \omega_h$ are used in the denominator.

Finally calculating the sensor output signal $\tilde{s}(\omega)$ using Eqn.~\eqref{eqn:sensor_output} considering the eigenmode decomposition~\eqref{eqn:v_th_decomposition} yields 

\begin{align}
\tilde{s}(\omega) = \sum_k \tilde{a}_k \, \tilde{g}_k
\end{align}
with the eigenmode dependent weighting factor
\begin{align}
\tilde{g}_k = \frac{\int \tilde{\boldsymbol{\varphi}}_k \cdot \hat{\mathbf{e}}_\text{ref} \; \text{d}\Omega_m}{\int \text{d}\Omega_m}
\end{align}

\subsection{FEM Discretization}
Finite-element discretization of Eqn.~\eqref{eqn:eigenvalue_problem_thermal} using the FEM discretized thermal field $\underline{\tilde{\mathbf{h}}}^\text{th}$ leads to
\begin{align} \label{eqn:eigenvalue_problem_thermal_FEM}
\omega \, \underline{B}_0 \, \underline{\tilde{\boldsymbol{\varphi}}}^\text{th}(\omega) = \underline{A}_{0\perp} \underline{\tilde{\boldsymbol{\varphi}}}^\text{th} - \gamma \, \underline{P}_0 \, \underline{C}^\text{ext} \, \underline{\tilde{\mathbf{h}}}^\text{th}(\omega),
\end{align}
with a weighted 3D mass matrix $\underline{C}^\text{ext} = \int J_s \Lambda_i \, \Lambda_j \, \text{d}\Omega_m$. The combined operator $\underline{M}_0 = \underline{P}_0 \, \underline{C}^\text{ext}$ can introduced which is equivalent to a weighted 2D mass matrix within the tangent-plane. The mode amplitudes can be derived using the same line of reasoning as for the continuum problem
\begin{align}
\tilde{a}_h = - \frac{\gamma \, \omega_h}{\omega - \omega_h - i \, \delta \omega_h} \, \left(\underline{\tilde{\boldsymbol{\varphi}}}_h, \underline{\tilde{\mathbf{h}}}^\text{th} \right)_{\underline{M}_0},
\end{align}
which directly leads the to following expression for the discrete transfer curve $\underline{H}(\omega)$:
\begin{align}
\tilde{s}(\omega) = \sum_k \tilde{a}_k \, \tilde{g}_k = \underbrace{-\sum_k \tilde{g}_k \, \frac{\gamma \, \omega_k}{\omega - \omega_k - i \, \delta \omega_k} \, \underline{\tilde{\boldsymbol{\varphi}}}^*_k \, \underline{M}_0}_{\underline{H}(\omega)} \, \underline{\tilde{\mathbf{h}}}^\text{th}
\end{align}

The discretized thermal field approximately yields the following autocorrelation
\begin{align}
R_{hh}(\tau) = \langle \underline{\mathbf{h}}_i^\text{th}(t) \; \underline{\mathbf{h}}_j^\text{th}(t') \rangle &\approx \frac{2 \alpha k_B T}{\gamma \overline{J_s}_i \overline{V}_i} \, \delta_{ij} \, \delta(t-t'),
\end{align}
where the nodal volume $\overline{V}_i$, which is one fourth of the volume of all adjacent elements, has been used. Note that the commonly used assumption of uncorrelated noise \cite{ragusa_full_2009} on each node is not strictly true when using a FEM discretization (see Appendix \ref{sec:FEM_discretization} for a more detailed description). The power spectral density of the discretized thermal noise field results in
\begin{align}
S_{hh}(\omega) &\approx \frac{2 \alpha k_B T}{\gamma \overline{J_s}_i \overline{V}_i} \, \delta_{ij}
\end{align}

Putting everything together the total PSD of the reads like
\begin{align}
S_{ss} = \sum_{h,k} \tilde{g}_h \frac{\gamma \, \omega_h}{\omega - \omega_h - i \, \delta \omega_h} \, \underline{\tilde{\boldsymbol{\varphi}}}^*_h \, \underline{M}_0 \, S_{hh}(\omega) \, \underline{M}_0 \, \underline{\tilde{\boldsymbol{\varphi}}}_k \, \frac{\gamma \, \omega_k}{\omega - \omega_k + i \, \delta \omega_k} \, \tilde{g}_k^*
\end{align}

\section{Numerical Experiments}
\begin{figure}[h!]
  \begin{minipage}{0.33\linewidth}
    \begin{overpic}[width=0.9\linewidth]{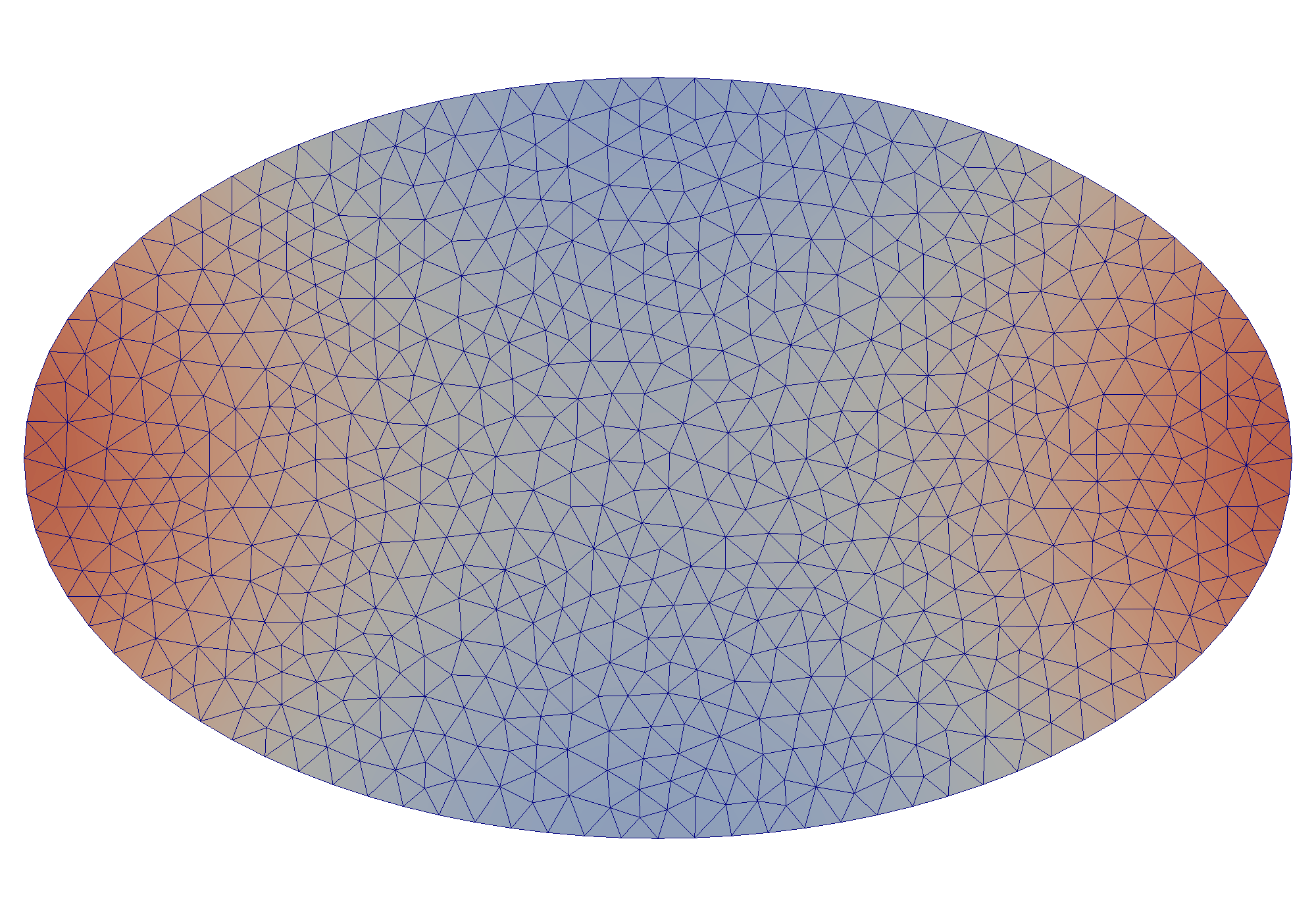}
      \put(20,-5){\small $f_1 = \SI{6.780}{\giga\hertz}$}
    \end{overpic}
  \end{minipage}%%
  \begin{minipage}{0.33\linewidth}
    \begin{overpic}[width=0.9\linewidth]{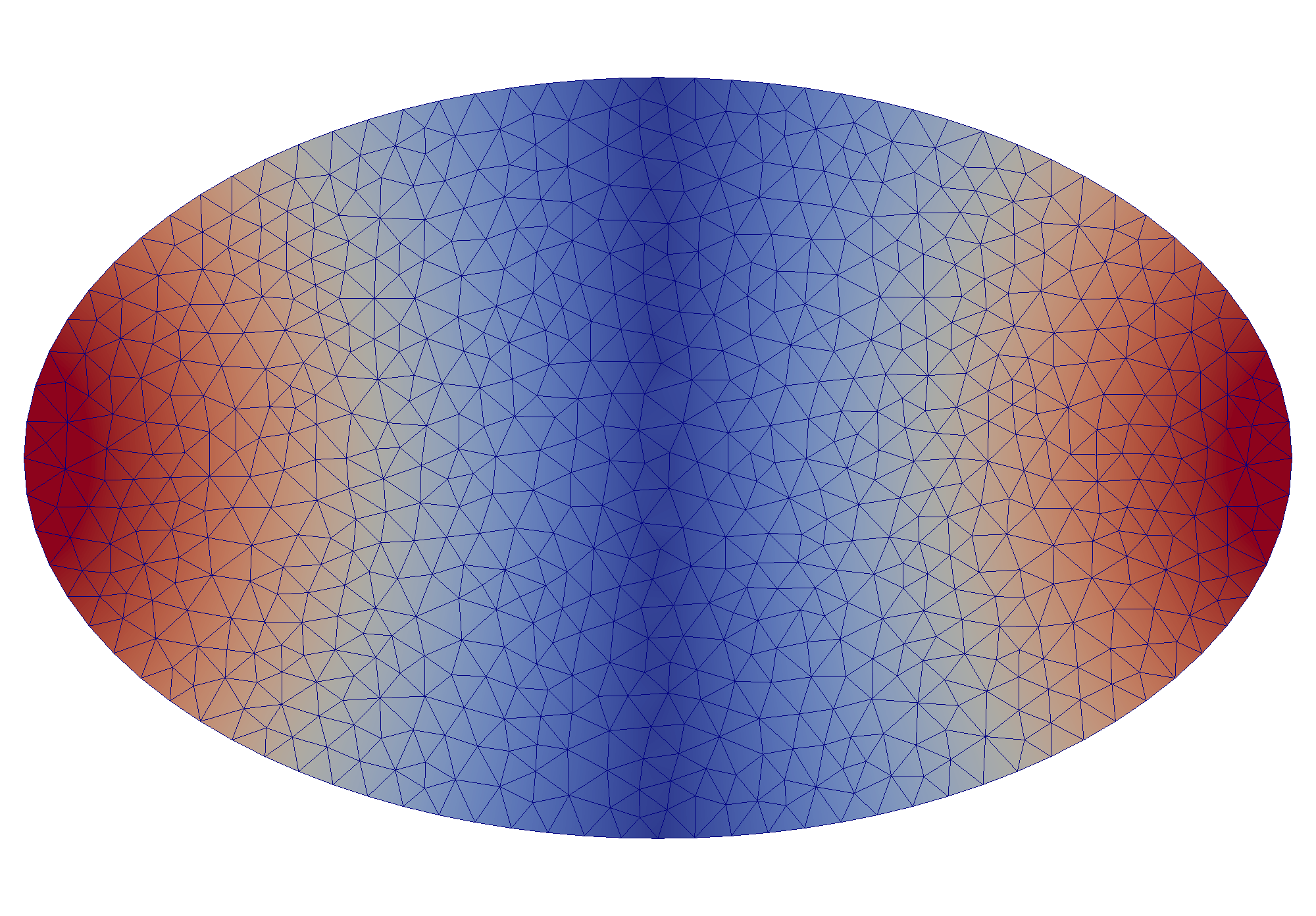}
      \put(20,-5){\small $f_2 = \SI{7.421}{\giga\hertz}$}
    \end{overpic}
  \end{minipage}%%
  \begin{minipage}{0.33\linewidth}
    \begin{overpic}[width=0.9\linewidth]{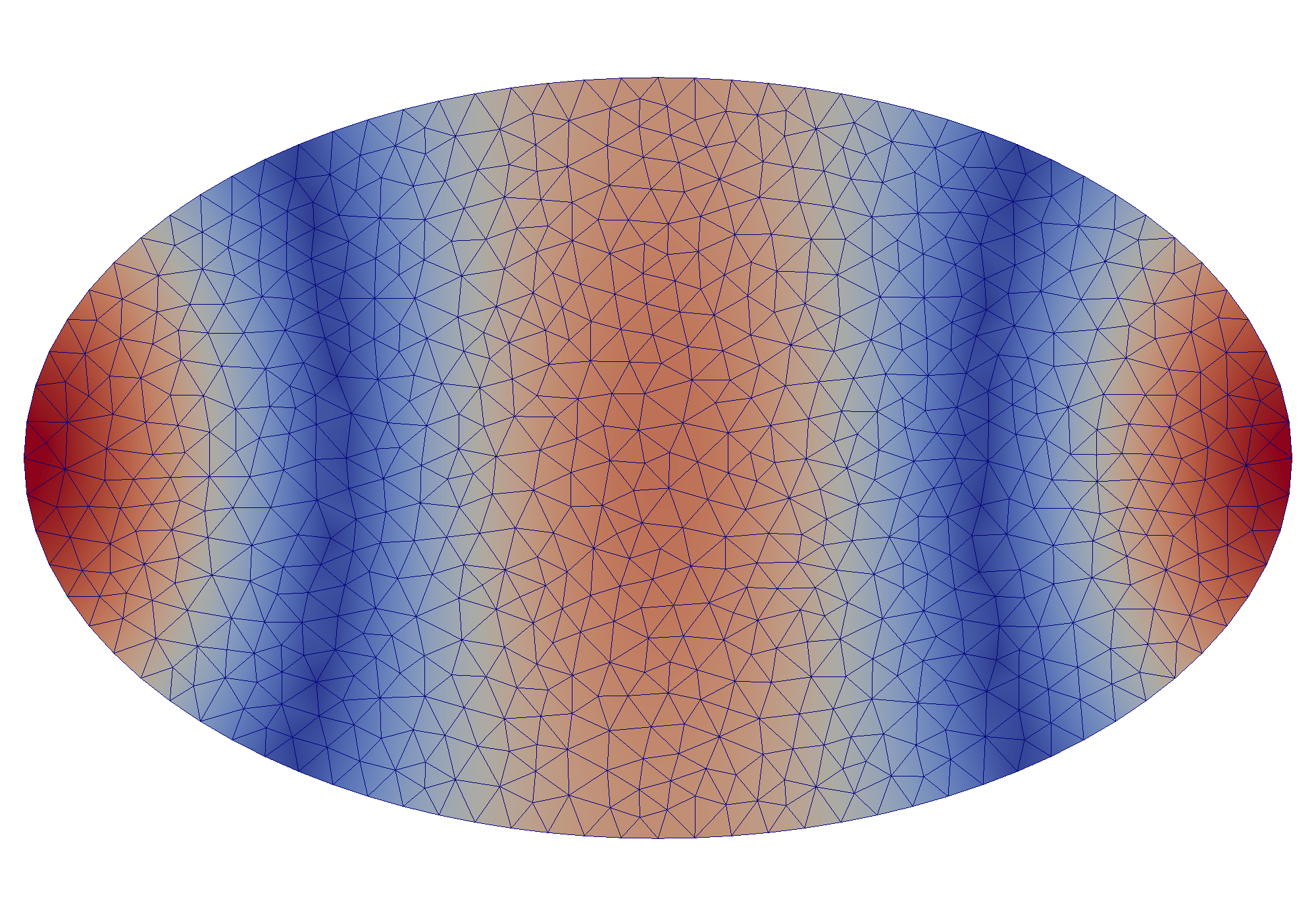}
      \put(20,-5){\small $f_3 = \SI{10.904}{\giga\hertz}$}
    \end{overpic}
  \end{minipage}%%
  \vspace{0.25cm}
  \begin{minipage}{0.33\linewidth}
    \begin{overpic}[width=0.9\linewidth]{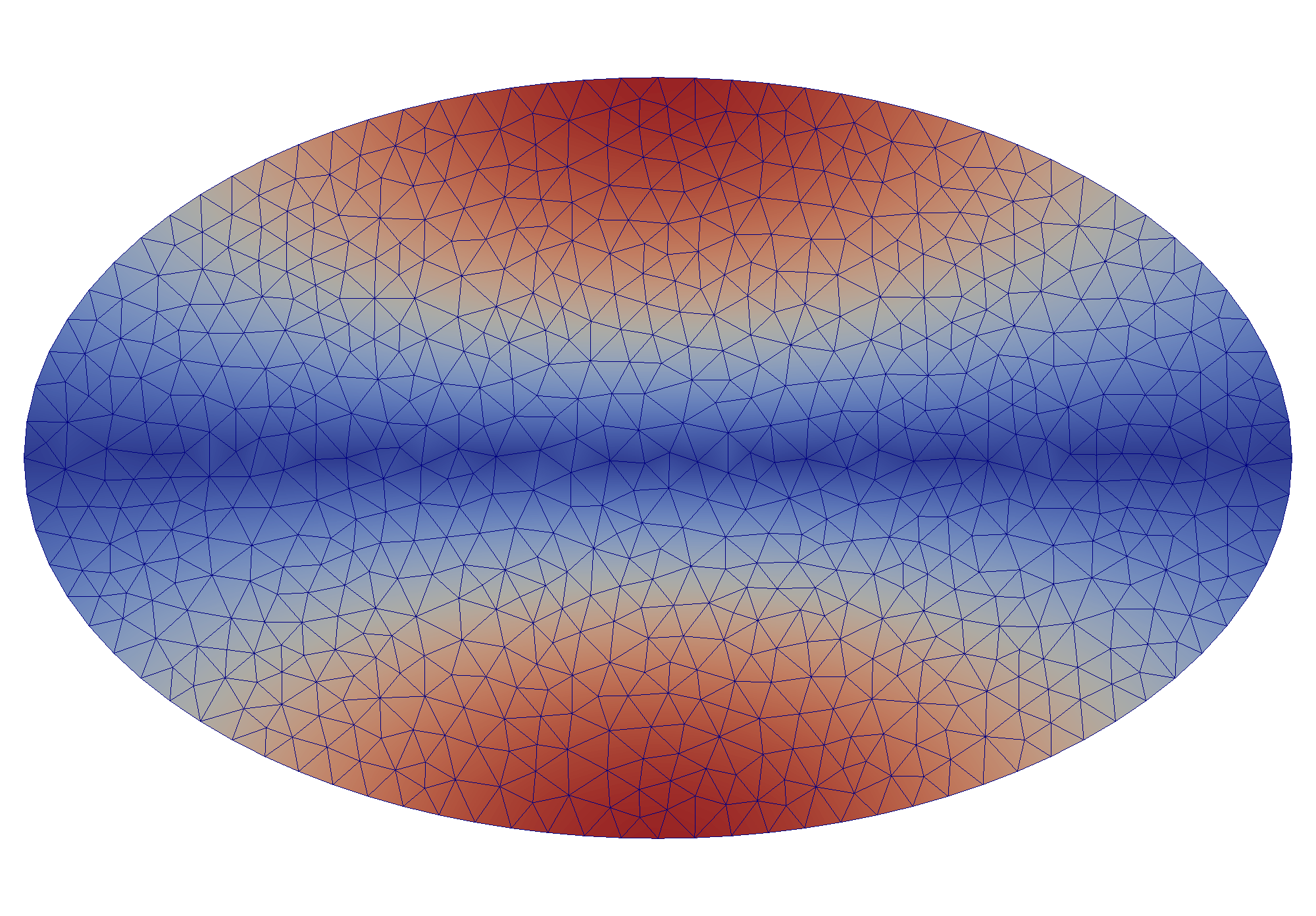}
      \put(20,-5){\small $f_4 = \SI{14.125}{\giga\hertz}$}
    \end{overpic}
  \end{minipage}%%
  \begin{minipage}{0.33\linewidth}
    \begin{overpic}[width=0.9\linewidth]{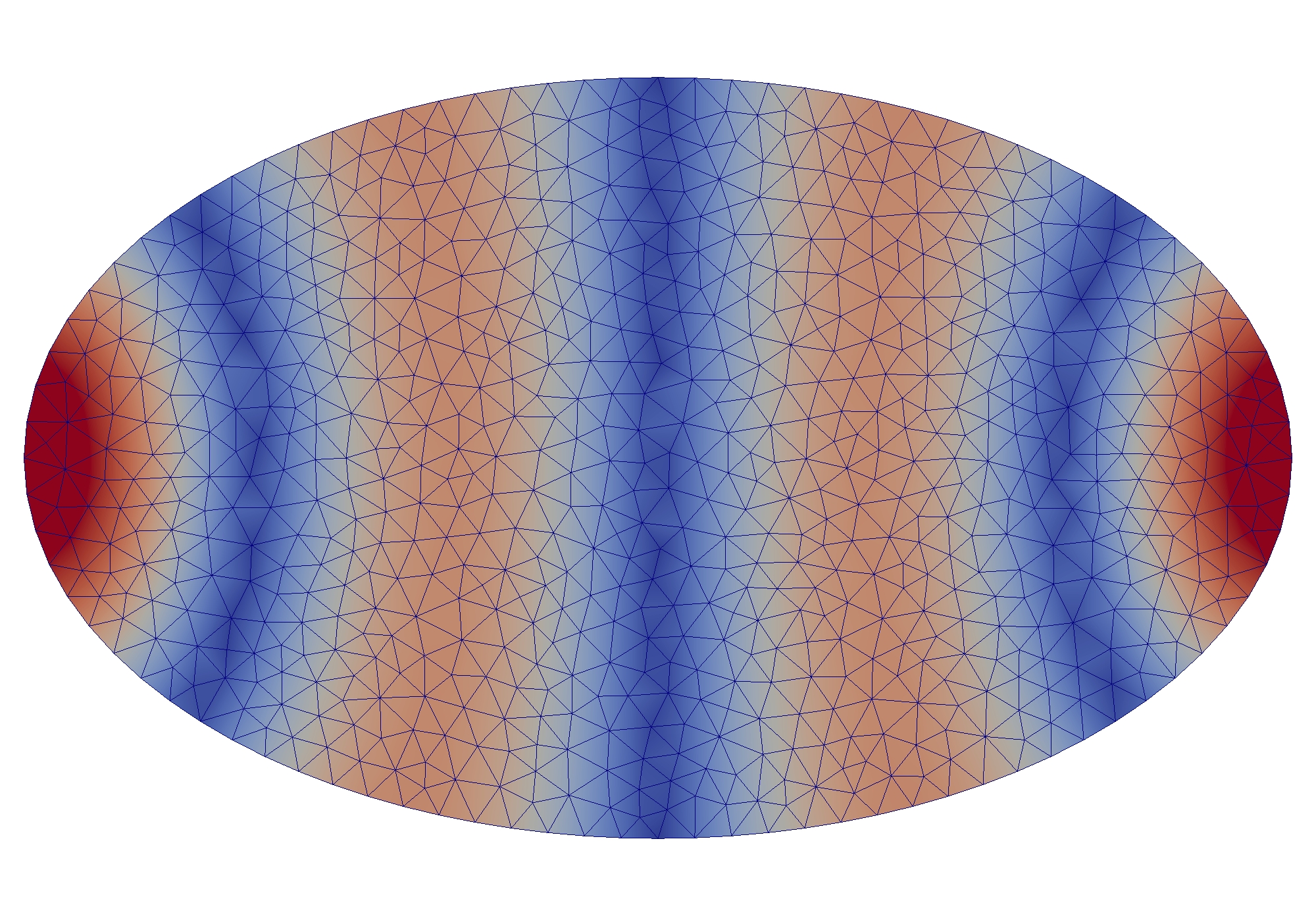}
      \put(20,-5){\small $f_5 = \SI{15.285}{\giga\hertz}$}
    \end{overpic}
  \end{minipage}%%
  \begin{minipage}{0.33\linewidth}
    \begin{overpic}[width=0.9\linewidth]{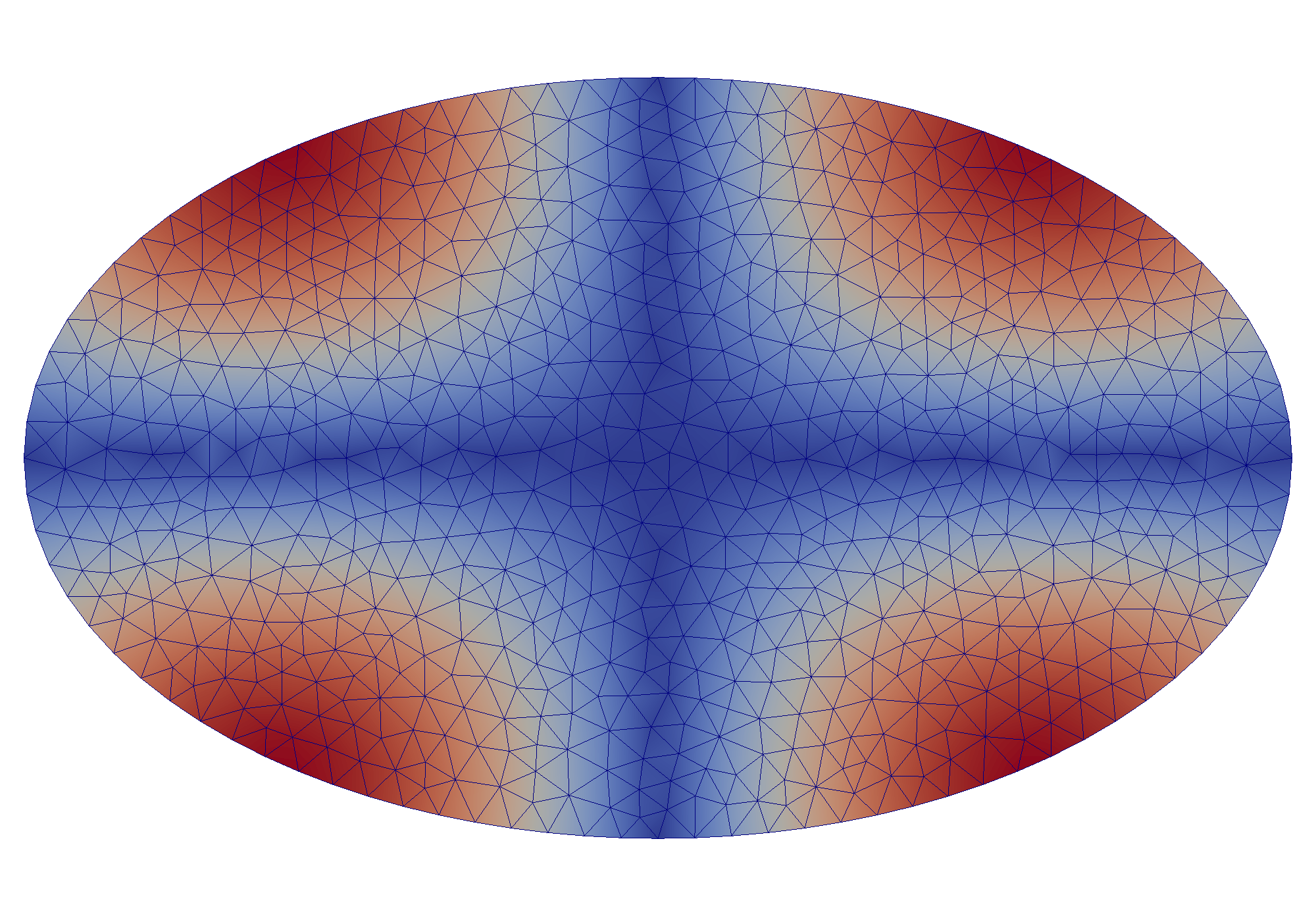}
      \put(20,-5){\small $f_6 = \SI{15.974}{\giga\hertz}$}
    \end{overpic}
  \end{minipage}%%
  \vspace{0.25cm}
  \begin{minipage}{0.33\linewidth}
    \begin{overpic}[width=0.9\linewidth]{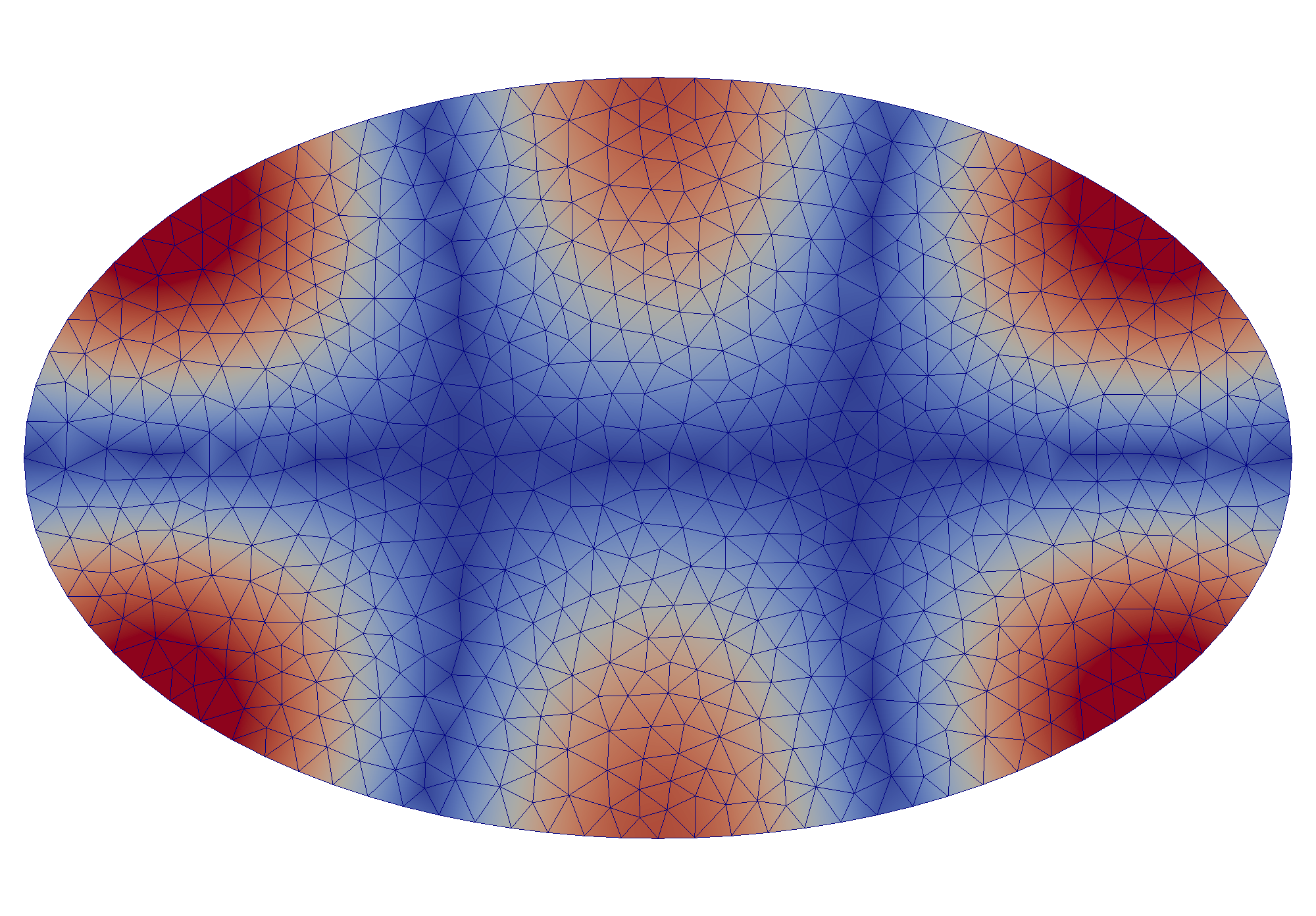}
      \put(20,-5){\small $f_7 = \SI{18.943}{\giga\hertz}$}
    \end{overpic}
  \end{minipage}%%
  \begin{minipage}{0.33\linewidth}
    \begin{overpic}[width=0.9\linewidth]{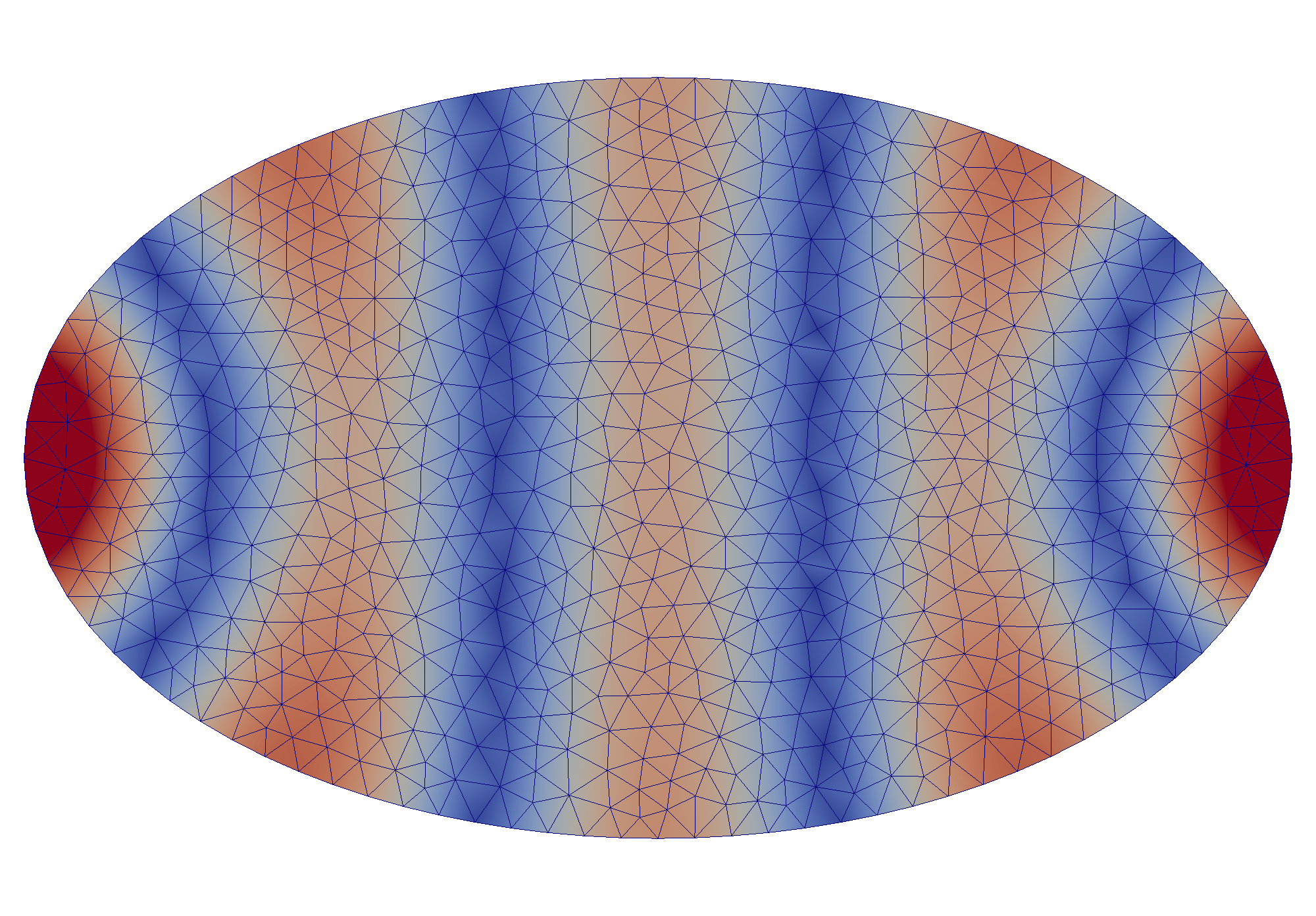}
      \put(20,-5){\small $f_8 = \SI{20.580}{\giga\hertz}$}
    \end{overpic}
  \end{minipage}%%
  \begin{minipage}{0.33\linewidth}
    \begin{overpic}[width=0.9\linewidth]{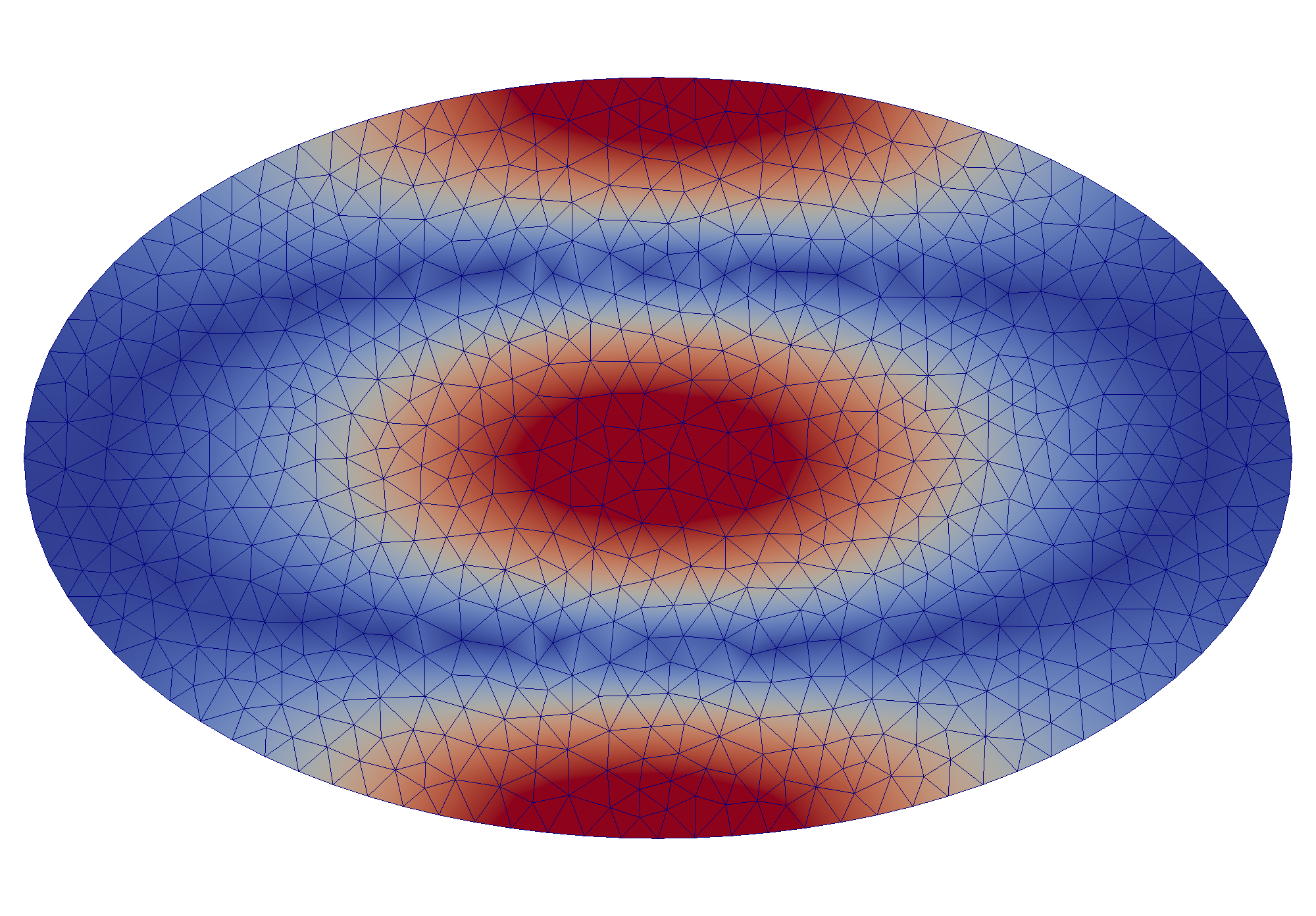}
      \put(20,-5){\small $f_9 = \SI{22.751}{\giga\hertz}$}
    \end{overpic}
  \end{minipage}%%
  \caption{Numerically computed oscillation amplitutes and frequencies of the first 9 eigenmodes of an elliptical $\SI{100}{\nano\metre} \times \SI{60}{\nano\metre} \times \SI{5}{\nano\metre}$ thin-film element using an average cell size of $\SI{3}{\nano\metre}$.}
  \label{fig:carlotti_eigenmodes}
\end{figure}

The proposed method for the calculation of resonant modes as well as the power spectral density has been applied to an elliptical nanodisc with a dimension of $\SI{100}{nm} \times \SI{60}{nm} \times \SI{5}{nm}$. The used micromagnetic material parameters are summarized in Tbl.~\ref{tbl:material}. By exciting the stationary system with a stochastic thermal noise field $\mathbf{h}^\text{th}$, and performing a node-wise Fourier transform, it is possible to numerically calculate the resonance frequencies as well as the corresponding eigenvectors \cite{carlotti_exchange_2014}. Results of the oscillation amplitudes as well as the corresponding resonance frequencies using the proposed method are visualized in Fig~\ref{fig:carlotti_eigenmodes}.

\begin{table}[h!]
  \centering
  \begin{tabular}{lcc}
  Quantity & Symbol & Value \\
  \hline
  saturation magnetization & $M_s$ & $\SI{860}{kA/m}$ \\
  exchange constant & $A$ & $\SI{13}{pJ/m}$ \\
  uniaxial anisotropy constant & $K_1$ & $\SI{10}{kJ/m^3}$ \\
  uniaxial easy axis & $\mathbf{e}_\text{u}$ & $(1, 0, 0)$ \\
  phenomenological damping constant & $\alpha$ & $0.02$ \\
  \end{tabular}
  \caption{Material parameters of the polycrystalline permalloy.}
  \label{tbl:material}
\end{table}

\begin{figure}[h!]
  \begin{tikzpicture}
    \begin{axis}[
      height=6cm,
      width=12cm,
      ytick={0,10,...,60},
      grid=major,
      legend pos=south east,
      xlabel={Index $n$, $m$ [1]},
      ylabel={Frequency [$\si{\giga\hertz}$]},
      ]

      % albert
      \addplot[color=red, thick, mark=x] coordinates{
        (0,  6.9556025370)
        (1,  7.3995771670)
        (2, 11.0993657505)
        (3, 15.6871035941)
        (4, 21.4587737844)
        (5, 27.9704016913)
        (6, 35.6659619450)
        (7, 44.3974630021)
        (8, 53.8689217759)
        (9, 64.5243128964)
      };
      \addplot[color=red, thick, mark=x, forget plot] coordinates{
        (0,  6.9556025370)
        (1, 14.7991543340)
        (2, 24.1226215645)
        (3, 37.1458773784)
        (4, 54.3128964059)
      };
      \addplot[color=red, thick, mark=x, forget plot] coordinates{
        (0,  6.9556025370)
        (1, 16.5750528541)
        (2, 33.2980972516)
        (3, 58.3086680761)
      };
      \addlegendentry{Albert \cite{albert_domain_2016}};

      % carlotti
      \addplot[color=green, thick, mark=+] coordinates{
        (0,  6.9556025370)
        (1,  7.3995771670)
        (2, 11.0993657505)
        (3, 15.6871035941)
        (4, 21.4587737844)
      };
      \addplot[color=green, thick, mark=+, forget plot] coordinates{
        (0,  6.9556025370)
        (1, 14.7991543340)
        (2, 24.1226215645)
        (3, 37.1458773784)
      };
      \addplot[color=green, thick, mark=+, forget plot] coordinates{
        (0,  6.9556025370)
        (1, 16.5750528541)
        (2, 33.2980972516)
      };
      \addlegendentry{Carlotti \cite{carlotti_exchange_2014}};

      % bruckner
      \addplot[color=blue, thick, mark=o] coordinates{
        (0, 6.801628380)
        (1, 7.465215110)
        (2, 10.99324152)
        (3, 15.49130129)
        (4, 20.96935313)
        (5, 27.53471395)
        (6, 35.04891477)
        (7, 43.67079729)
        (8, 53.42645593)
      };
      \addplot[color=blue, thick, mark=o, forget plot] coordinates{
        (0, 6.801628380)
        (1, 14.23816784)
        (2, 23.18778818)
        (3, 35.72875930)
        (4, 52.68607175)
      };
      \addplot[color=blue, thick, mark=o, forget plot] coordinates{
        (0, 6.801628380)
        (1, 16.17099912)
        (2, 32.36849512)
      };
      \addlegendentry{Bruckner et al.};

      \draw (axis cs:8,53.4) node[below right] {$(n,0)$};
      \draw (axis cs:4,52.7) node[below right] {$(0,n)$};
      \draw (axis cs:2,32.4) node[above left ] {$(n,n)$};
    \end{axis}
  \end{tikzpicture}
  \caption{Comparison of the numerically computed oscillation frequencies of an elliptical $\SI{100}{\nano\metre} \times \SI{60}{\nano\metre} \times \SI{5}{\nano\metre}$ thin-film element using an average cell size of $\SI{3}{\nano\metre}$. The eigenmodes are aranged in three different branches according to their node indices $(n,m)$.}
  \label{fig:carlotti_frequencies}
\end{figure}
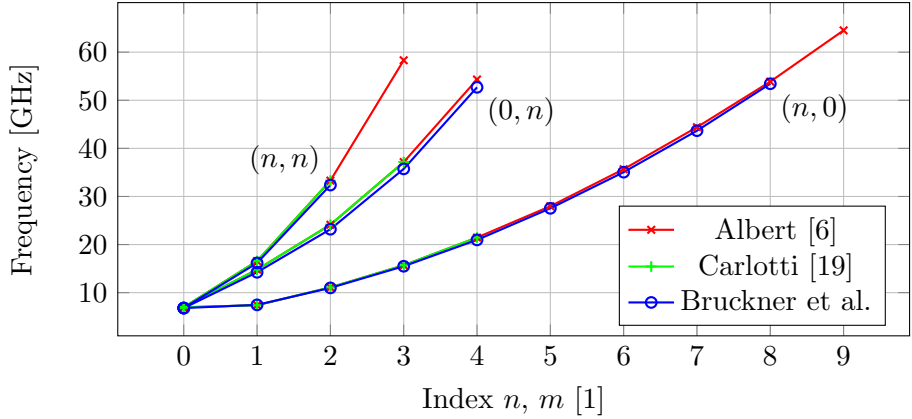

The original resonance frequencies calculated from stochastic time-integration \cite{carlotti_exchange_2014} are in perfect agreement with the ones calculated by Albert using an eigenmode based approach \cite{albert_domain_2016,albert_frequency_2016}. A comparison of the different methods is presented in Fig.~\ref{fig:carlotti_frequencies}.

Finally the PSD is calculated using the proposed semi-analytical method. For sake of comparison a stochastic time-integration of the same system is performed using a time-step $\Delta t = \SI{10}{fs}$ and a total simulation time of $T=\SI{40}{ns}$, which leads to a frequency resolution $\Delta f = 1/T = \SI{25}{MHz}$ and a total number of samples $N=T/\Delta t = \SI{4e6}{}$. The PSD can be calculated numerically by using the discrete form of Eqn.~\eqref{eqn:PSD}:
\begin{align} \label{eqn:PSD_discrete}
S_{xx}(\omega) \approx \frac{1}{T} \left\vert \sum_{n=1}^N x_n \, e^{-i \omega t_n} \, \Delta t \right\vert^2,
\end{align}
where $x(t)$ is the average magnetization in the reference direction $\mathbf{e}^\text{ref}$ and $x_n = x(t_n)$ and $t_n = n \, \Delta t$. The fast Fourier transform (FFT) can be used for efficient summation. In this case the result will only be calculated for discrete angular frequencies $w_k = 2 \pi f_k$, with $f_k = \frac{k}{N\, \Delta t}$ and integer $k = -\frac{N}{2} \dots \frac{N}{2}-1$ (for even $N$).

\begin{figure}[h!]
  \includegraphics[width=1\columnwidth]{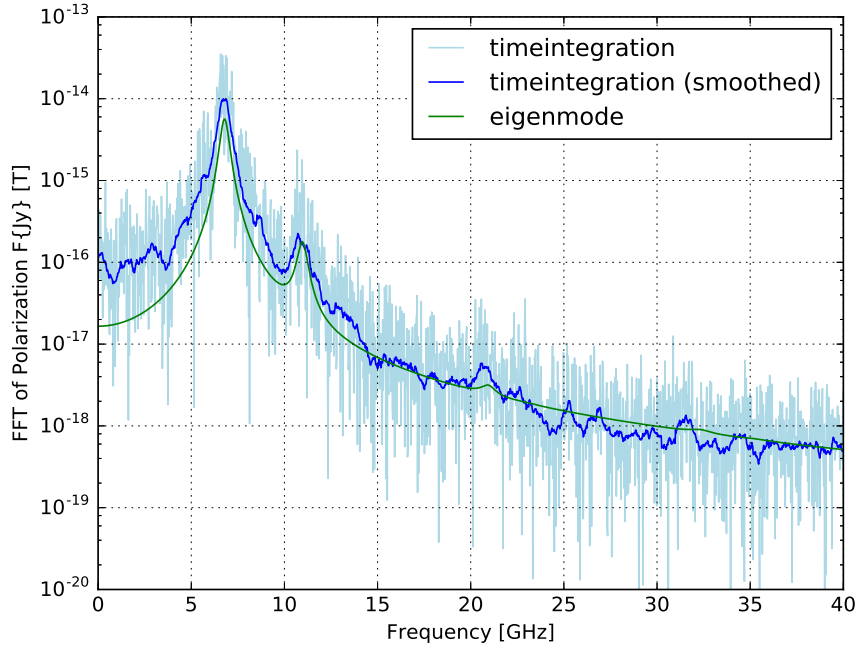}
  \caption{Comparison of the PSD calculation of an elliptical $\SI{100}{\nano\metre} \times \SI{60}{\nano\metre} \times \SI{5}{\nano\metre}$ thin-film element using an average cell size of $\SI{3}{\nano\metre}$ and a temperature of $T=\SI{1}{K}$, by using (a) numerical transformation of the time-signal of the stochastic integration, (b) moving average of the numerical transformation using a 30 samples avarage, (c) semi-analytic calculation using the presented eigenmode approach.}
  \label{fig:carlotti_PSD}
\end{figure}

The numerically calculated PSD is compared with the semi-analytical eigenmode results in Fig.~\ref{fig:carlotti_PSD}. Since the numerical results are very noisy, additionally a moving average over 30 samples of the raw data is calculated. Results are in good agreement, but the eigenmode based results are much more smooth. Additionally the calculation time could be reduced from $\SI{2}{days}$ (for the stochastic time-integration) to $\SI{300}{s}$ (for the eigenmode-based method).

\appendix
\section{Linearized LLG} \label{sec:linearized_LLG}
The following alternative form of the LLG (see e.g. \cite{alouges_new_2008}) is used for the linearization
\begin{align}
-\alpha \, \dot{\mathbf{m}} - \mathbf{m} \times \dot{\mathbf{m}} &= -\gamma \, \mathbf{h}^\text{eff}[\mathbf{m}] + \gamma \, \left( \mathbf{m} \cdot \mathbf{h}^\text{eff}[\mathbf{m}] \right) \; \mathbf{m},
\end{align}
which directly results in a generalized eigenvalue problem. Alternatively one could start with the Gilbert form of the LLG resulting in a standard eigenvalue problem, but this would complicate the FEM discretization since the total system matrix is not Hermitian \cite{daquino_novel_2009}.

Setting $\mathbf{m} = \mathbf{m}_0 + \varepsilon \, \mathbf{v}$ into the alternative form of the LLG equation, where $\mathbf{m}_0$ is a stable equilibrium configuration, and considering only terms linear in $\varepsilon$ yields
\begin{align}
\begin{split}
-\alpha \, \dot{\mathbf{v}} - \mathbf{m}_0 \times \dot{\mathbf{v}} &= -\gamma \, \mathbf{h}^\text{lin}[\mathbf{v}] + \gamma \, \left[ (\underbrace{\mathbf{v} \cdot \mathbf{h}^\text{eff}[\mathbf{m}_0]}_{= 0}) \, \mathbf{m}_0 + (\mathbf{m}_0 \cdot \mathbf{h}^\text{lin}[\mathbf{v}]) \, \mathbf{m}_0 + (\underbrace{\mathbf{m}_0 \cdot \mathbf{h}^\text{eff}[\mathbf{m}_0]}_{= h_0}) \, \mathbf{v} \right] \\
&= \left( \mathbb{1} - \mathbf{m}_0 \otimes \mathbf{m}_0 \right) \, \gamma \, \left( h_0 \, \mathbb{1} - \mathbf{h}^\text{lin} \right)[\mathbf{v}] \\
&= \mathcal{P}_0 \, \mathcal{A}_0[\mathbf{v}],
\end{split}
\end{align}
where $\mathbf{h}^\text{lin}$ denotes the part of the effective field $\mathbf{h}^\text{eff}[\mathbf{m}] = \mathbf{h}^\text{lin}[\mathbf{m}] + \mathbf{h}^\text{ext}$, which is linear in $\mathbf{m}$. The first term within the square brackets vanishes, since the fluctuation $\mathbf{v} \perp \mathbf{m}_0$ due to the normalization constraint, and the effective field $\mathbf{h}^\text{eff}[\mathbf{m}_0] \parallel \mathbf{m}_0$ due to Brown's equation and the fact the $\mathbf{m}_0$ represents an equilibrium configuration. The third term within the square brackets describes the influence of the parallel component of the effective field at equilibrium, and is orthogonal to the equilibrium magnetization $\mathbf{m}_0$. Thus the tangent-plane projection operator $\mathcal{P}_0$ introduced in the second line can be extended over both remaining terms.

\section{FEM discretization of effective field} \label{sec:FEM_discretization}
The micromagnetic energy contains the following contributions
\begin{align}
\mathcal{E} &= \mathcal{E}^\text{ex} + \mathcal{E}^\text{ani} + \mathcal{E}^\text{d} + \mathcal{E}^\text{ext} + \dots &
\end{align}
with the exchange energy $E^\text{ex}$, the uniaxial anisotropy energy $E^\text{ani}$, the magnetostatic energy $E^\text{d}$, and the Zeeman energy $E^\text{ext}$. Additionally terms can easily be added to the formulation, however a linearization is necessary in case of non-linear interactions like e.g. the cubical anisotropy.

The energy can also be expressed in terms of an explicit effective field $\mathbf{h}^\text{eff}[\mathbf{m}] = \mathbf{h}^\text{lin}[\mathbf{m}] + \mathbf{h}^\text{ext}$, which yields
\begin{align}
\mathcal{E} &= -\frac{1}{2} \int J_s \, \mathbf{m} \cdot \mathbf{h}^\text{lin}[\mathbf{m}] \, \text{d}\Omega_\text{m} - \int J_s \, \mathbf{m} \cdot \mathbf{h}^\text{ext} \, \text{d}\Omega_\text{m}
\end{align}
Starting from the energy, the effective field can be defined by means of the functional derivative
\begin{align}
\frac{\delta \mathcal{E}}{\delta \mathbf{m}} &= -J_s \, \mathbf{h}^\text{lin}[\mathbf{m}] - J_s \, \mathbf{h}^\text{ext} = -J_s \, \mathbf{h}^\text{eff}[\mathbf{m}]
\end{align}

The continuous expressions for the micromagnetic energies and the corresponding fields are given as
\begin{align}
\mathcal{E}^\text{ex} &= \int A_\text{ex} (\boldsymbol{\nabla} \mathbf{m})^2 \, \text{d}\Omega_\text{m} &
\mathbf{h}^\text{ex}[\mathbf{m}] &= \frac{2 A_\text{ex}}{J_s} \, \Delta \mathbf{m} \\
\mathcal{E}^\text{ani} &= -\int K_\text{u} (\mathbf{m} \cdot \mathbf{e}_\text{u})^2 \, \text{d}\Omega_\text{m} &
\mathbf{h}^\text{ani}[\mathbf{m}] &= \frac{2 K_\text{u}}{J_s} \, \mathbf{e}_\text{u} (\mathbf{e}_\text{u} \cdot \mathbf{m}) \\
\mathcal{E}^\text{d} &= \frac{1}{2} \int J_s \, \mathbf{m} \cdot \boldsymbol{\nabla} u \, \text{d}\Omega_\text{m} &
\mathbf{h}^\text{d}[\mathbf{m}] &= -\boldsymbol{\nabla} u(\mathbf{m}) \\
\mathcal{E}^\text{ext} &= - \int J_s \, \mathbf{m} \cdot \mathbf{h}^\text{ext} \, \text{d}\Omega_\text{m}
\end{align}
with exchange constant $A_\text{ex}$, uniaxial anisotropy constant $K_\text{u}$, easy axis direction $\mathbf{e}_\text{u}$ and saturation polarization $J_s$. The magnetic scalar potential $u$ is defined by
\begin{align} \label{eqn:magnetic_scalar_potential}
\begin{split}
&\Delta u = \frac{1}{\mu_0}\boldsymbol{\nabla} \cdot (J_s \, \mathbf{m}) \quad \text{in} \; \mathbb{R}^3 \\
&u\vert_{r \rightarrow \infty} \rightarrow 0
\end{split}
\end{align}

Directly discretizing the exchange field using a Galerkin approach and Lagrange $\mathcal{P}_1$ elements, leads to problems with the partial integration needed in order to avoid the second derivative. Considering two different materials, where $\frac{2A_\text{ex}}{J_s}$ is not continuous, leads to non-vanishing boundary conditions. The problem can be avoided by instead discretizing $J_s \, \mathbf{h}^\text{ex}$, where it is known from micromagnetic theory that the occuring boundary term $A_\text{ex} \, (\mathbf{n} \cdot \nabla \mathbf{m})$ is continuous and therefor the boundary integral vanishes.

Additionally the discretization of $J_s \, \mathbf{h}^\text{eff}$ directly yields a discrete form of the total energy
\begin{align}
E = -\frac{1}{2} \, \underline{\mathbf{m}} \, \underline{C}^\text{lin} \, \underline{\mathbf{m}} - \underline{\mathbf{m}} \, \underline{C}^\text{ext} \, \underline{\mathbf{h}}^\text{ext}
\end{align}
with the following discrete matrix representations of the corresponding field operators
\begin{align}
\underline{\mathbf{C}}^\text{ex} &= \int 2 A_\text{ex} \, \nabla \Lambda_i \cdot \nabla \Lambda_j \, \text{d}\Omega_\text{m} \\
\underline{\mathbf{C}}^\text{ani} &= -\int 2 K_\text{u} \, (\Lambda_i \cdot \mathbf{e}^\text{u}) \, (\Lambda_j \cdot \mathbf{e}^\text{u}) \, \text{d}\Omega_\text{m} \\
\underline{\mathbf{C}}^\text{ext} &= \int J_s \, \Lambda_i \, \Lambda_j \, \text{d}\Omega_\text{m}
\end{align}

The discretization of the demagnetization field operator $\underline{\mathbf{C}}^\text{d}$ requires the discrete solution of the magnetic scalar potential $u$ in Eqn.~\eqref{eqn:magnetic_scalar_potential}. Using a FEM only approach requires to discretize a (large enough) airbox around the magnetic region, since the boundary conditions for $u$ is only known at infinity. The commonly used Fredkin-Koehler method \cite{fredkin_hybrid_1990} avoids this problem by using the boundary element method (BEM) in combination with a FEM discretization of the magnetic domain. Note that the Fredkin-Koehler method is not perfectly symmetric, which leads to small non-Hermitian contribution of the discretized effective field operator. However numerical tests using non-Hermitian eigenvalue solvers show that for typical problems the resulting effects are negligible.

Generally the discrete linear field operators as used in Eqn.~\eqref{eqn:FEM_discretization} can be expressed as
\begin{align}
\underline{\mathbf{C}}^\text{lin} &= \int J_s \, \Lambda_i \, \mathbf{h}^\text{lin}[\Lambda_j] \, \text{d}\Omega_\text{m}
\end{align}

The stochastic thermal field is a special case of an external field, since it does not depend on $\mathbf{m}$. For the semi-analytical calculation of the PSD the discretized autocorrelation ot the thermal noise $\mathbf{h}^\text{th}$ is needed. The Galerkin discretization of the termal field can be written as
\begin{align}
x^\text{th}_i = \underbrace{\int J_s \, \Lambda_i \, \Lambda_j \, \text{d}\Omega_\text{m}}_{\underline{C}^\text{ext}} \, h^\text{th}_j = \int J_s \, \Lambda_i \, \mathbf{h}^\text{th}(\mathbf{x}) \, \text{d}\Omega_\text{m},
\end{align}
where again both sides are multiplied with $J_s$ and the quantity $x_i$, which also occurs as source term of the discrete, inhomogeneous eigenvalue equation \eqref{eqn:eigenvalue_problem_thermal_FEM}, is introduced. The discretized autocorrelation of $x_i$ results in
\begin{align} \label{eqn:autocorrelation_xx}
\begin{split}
R_{xx} = \left\langle x^\text{th}_i \, x^\text{th}_j \right\rangle &= \left\langle \int J_s \, \Lambda_i \, \mathbf{h}^\text{th}(\mathbf{x}) \, \text{d}\Omega_\text{m} \; \int J_s \, \Lambda_j \, \mathbf{h}^\text{th}(\mathbf{x'}) \, \text{d}\Omega'_\text{m} \right\rangle = \\
&= \iint \text{d}\Omega_\text{m} \, \text{d}\Omega'_\text{m} \, J_s^2 \, \Lambda_i \, \Lambda_j \, \underbrace{\left\langle \mathbf{h}^\text{th}(\mathbf{x}) \mathbf{h}^\text{th}(\mathbf{x'}) \right\rangle}_{\frac{2 \alpha k_B T}{\gamma J_s} \, \delta(\mathbf{x}-\mathbf{x}') \, \delta(t-t') } = \\
&= \frac{2 \alpha k_B T}{\gamma} \, \int J_s \, \Lambda_i \, \Lambda_j \, \text{d}\Omega_\text{m} \; \delta(t-t'),
\end{split}
\end{align}
which shows that the discretized noise on next-neighbor nodes is not uncorrelated. Nevertheless, for the stochastic time-integration, where the discrete thermal field $h_i^\text{th}$ is explicitly needed, one often expresses the autocorrelation of $h_i^\text{th}$ by using a mass-lumping approximation
\begin{align}
\int J_s \, \Lambda_i \, \Lambda_j \, \text{d}\Omega_\text{m} \approx \overline{V}_i \, \overline{J_s}_i\, \delta_{ij}
\end{align}
with the average polarization $\overline{J_s}_i$ and the average volume $\overline{V}_i$ at node $i$. Averageing is performed over all adjecent tetrahedra $\Delta_k$ using the cell volume $V_k$ and constant cell polarization ${J_s}_k$)
\begin{align}
\overline{J_s}_i = \frac{\sum_{\Delta_k} V_k \, {J_s}_k}{\sum_{\Delta_k} V_k}
& &
\overline{V}_i = \frac{1}{4} \sum_{\Delta_k} V_k
\end{align}
Utilizing the diagonality of the lumped operator allows to express the autocorrelation of the discrete thermal field as
\begin{align}
\begin{split}
R_{hh} = \left\langle h^\text{th}_i \, h^\text{th}_j \right\rangle &\approx \overline{V}_i^{-1} \, \overline{J_s}_i^{-1} \, \left\langle x^\text{th}_i \, x^\text{th}_j \right\rangle \, \overline{V}_i^{-1} \, \overline{J_s}_i^{-1} \\
&= \frac{2 \alpha k_B T}{\gamma \overline{V}_i \overline{J_s}_i} \delta_{ij} \, \delta(t-t'),
\end{split}
\end{align}
which appoximately represents spatially uncorrelated, as it is commonly used in various time-integration codes \cite{ragusa_full_2009,scholz_micromagnetic_2001,mentink_stable_2010}. Note that in Eqn.~\eqref{eqn:autocorrelation_xx} constant damping $\alpha$ and temperature $T$ is assumed. If those quantities vary in space the corresponding lumped quantities $\overline{\alpha}_i$ and $\overline{T}_i$ need to be used, instead.

%\section{Alternative derivation of PSD without orthogonality relation}

%\section{Stoner Wohlfarth Example}
%For clarification of the method the calculation of the PSD should be demonstrated for a domain Stoner-Wohlfarth particle, which can be represented by a single magnetic moment. For sake of simplicity one assumes
%
%TODO:
%\begin{itemize}
%  \item $a_k$ use tilde?
%\end{itemize}

%\cite{smith}
%\cite{Dynamics of skyrmionic states in confined helimagnetic nanostructure}
%\cite{Thermal stability of metastable magnetic skyrmions: Entropic narrowing and significance of internal eigenmodes}

\section{Acknowledgements}
The financial support by the Austrian Federal Ministry for Digital and Economic Affairs and the National Foundation for Research, Technology and Development is gratefully acknowledged.

\bibliographystyle{ieeetr}
\bibliography{refs}

\end{document}